\documentclass[nato,noid,numreferences]{crckbked}

\usepackage{graphicx}
\usepackage{epsf}
\begin{document}

\begin{opening}
\title{TOWARDS THE TOPOLOGICAL SUSCEPTIBILITY WITH OVERLAP FERMIONS}



\author{Tam\'as G.\ Kov\'acs$^\dagger$}
\institute{NIC/DESY, Platanenallee 6\\
           D-15738 Zeuthen, Germany }

\begin{abstract}
Using a reweighting technique combined with a low-mode 
truncation of the fermionic determinant, we estimate
the quark-mass dependence of the QCD topological susceptibility 
with overlap fermions. In contrast to previous lattice simulations
which all used non-chiral fermions,
our results appear to be consistent with the simple continuum model
of D\"urr. This indicates that at current lattice spacings 
the use of the index theorem might not be justified and
the fermionic definition of the charge might be needed. 
\end{abstract}

\end{opening}

%
%
\renewcommand{\thefootnote}{\fnsymbol{footnote}}
\footnotetext[0]{$^\dagger$On leave from the Department of 
Theoretical Physics, University of P\'ecs, Hungary. 
Supported by the EU's Human Potential 
Program under contract HPRN-CT-2000-00145, and by Hungarian 
science grant OTKA-T032501.}
\renewcommand{\thefootnote}{\arabic{footnote}}

Smooth $SU(3)$ gauge field configurations can be characterised 
with an integer charge $Q$, a topological invariant.
Topology plays a crucial role in shaping the low-energy 
behavior of QCD. Instantons are believed to be largely 
responsible for the spontaneous breaking of chiral symmetry.
The axial anomaly \cite{'tHooft:1986nc} as well as the large 
mass of the $\eta'$ \cite{Witten:1979vv,Veneziano:1979ec} are also 
intimately connected to gauge configurations of non-trivial
topology. For a good understanding of QCD, it is thus
crucial to have a consistent picture of its topological
structure.

Fluctuations of the topological charge can be characterised 
with the topological susceptibility,
\begin{equation}
 \chi = \frac{\langle Q^2 \rangle}{V},
\end{equation}
where $\langle . \rangle$ denotes averaging with respect to the
full QCD measure and $V$ is the volume of the system.
One of the most profound effects that light dynamical fermions 
are expected to have on the QCD vacuum is the suppression of
fluctuations of the topological charge. Chiral 
perturbation theory predicts that in the presence of $N_f$ 
degenerate light fermion flavours the susceptibility vanishes
with the quark mass as 
\begin{equation}
 \lim_{m\rightarrow 0} \chi \, = \, \frac{\Sigma m}{N_f^2} + {\cal O}(m^2)
        \, = \, \frac{f_\pi^2 m_\pi^2}{2N_f} + {\cal O}(m_\pi^4),
     \label{eq:LS}
\end{equation}
where $m_\pi$ and $f_\pi$ are the pion mass and decay constant,
$\Sigma$ is the chiral condensate and $m$ is the quark mass
\cite{Leutwyler:1992yt}. 

At the other extreme, when the quarks are very heavy, 
their influence eventually becomes negligible and the 
susceptibility approaches its quenched value
which is known to be $\chi_q=(203\pm 5$MeV$)^4$ 
\cite{Hasenfratz:1998qk}. Recently it has also been
discussed by S.\ D\"urr how the susceptibility is 
expected to behave for intermediate quark masses 
\cite{Durr:2001ty}. Starting with the observation that
besides the light quarks, the unit volume\footnote{Recall that
the susceptibility is defined as the fluctuation of $Q$ {\it per 
unit volume}.}
also suppresses higher topological sectors and noting that
these two mechanisms are independent, he derived a phenomenological
formula for the susceptibility. In terms of the
pion mass it reads as 
\begin{equation}
 \chi(m_\pi) = \left( \frac{2 N_f}{m_\pi^2 f_\pi^2} +
                      \frac{1}{\chi_q} \right)^{-1}.
     \label{eq:Durr}
\end{equation}
Both $f_\pi$ and $\chi_q$ are independently known therefore
this formula has no free parameter. While (\ref{eq:Durr})
does not come from first principles, it is based on a 
physically appealing plausible picture. Even if it is not the
last word in this subject, it indicates that the susceptibility
might substantially deviate from the Leutwyler-Smilga curve
already at smaller quark masses than previously expected
and it might be strongly suppressed even at moderately
heavy quark masses.

Although it is only the quenched topological susceptibility
that has an immediate phenomenological interest, being 
connected to the $\eta'$ mass through the Witten-Veneziano
formula \cite{Witten:1979vv,Veneziano:1979ec}, it would
also be of considerable interest to check the expected chiral 
behaviour of the full (unquenched) susceptibility. This is because
in the derivation of the Witten-Veneziano formula it is
always tacitly assumed that the full susceptibility goes
to zero in the chiral limit (see e.g. \cite{Giusti:2001xh}
for a recent discussion on the lattice). 

Recently several attempts have been made to check 
these predictions against numerical simulations on the lattice but 
the situation is still rather controversial. 
Some lattice simulations show only very slight or no suppression of
topological fluctuations \cite{AliKhan:2001ym,Bali:2001gk,Alles:2000cg}. 
The latest UKQCD results \cite{Hart:2001pj} obtained with
improved Wilson fermions agree well with the Leutwyler-Smilga 
prediction of Eq.\ (\ref{eq:LS}), even beyond its expected range
of validity and the UKQCD susceptibility is still significantly higher
than what D\"urr's interpolation formula (\ref{eq:Durr}) anticipates.

To understand why lattice simulations do not show the 
expected fermionic suppression of the topological susceptibility,
let us briefly look at the physical mechanism that
leads to this suppression. In the background of a charge $Q$ 
smooth gauge configuration the Dirac operator $D$ has (at least)
$|Q|$ zero eigenvalues for each flavour, due to the Atiyah-Singer 
index theorem. The quark effective action --- obtained after integrating 
out the fermions --- for $N_f$ flavours is proportional to det$^{N_f}(D+m)$.
In the light quark limit the determinant is $\propto m^{|Q|N_f}$
which leads to the suppression of higher $Q$ 
sectors and thus the topological susceptibility.
All this is true for smooth gauge field configurations in the 
continuum. The situation in lattice simulations however is rather
different since lattice Dirac operators in general do not have exact zero
eigenvalues and it is not obvious how the index theorem is 
realised on the lattice.

For a more thorough understanding of what happens on the
lattice, we first note that both the Witten-Veneziano formula
and the Leutwyler-Smilga relation follow from the flavour
singlet axial Ward Identity (WI). A general (Wilson type) lattice fermion 
action can be written as 
\begin{equation}
 S(\theta) = \overline{\psi} 
      \left[ D^A + e^{i\gamma_5\theta}(W + m_cM) +
                                      (m-m_c)M \right]\psi,
     \label{eq:Daction}
\end{equation}
where $D^A$ is the naive lattice Dirac operator, $W$ is the
Wilson term, $M$ the mass term and we already split off
the critical mass $m_c$ to account for the non-trivial 
mixing between the Wilson and the mass term, and the
$\theta$-dependence is also included \cite{Bochicchio:1984hi}.
Using that $D^A$ anticommutes, while the rest of the 
Dirac operator commutes with $\gamma_5$, it is easy to show
that the $\theta$-dependence can be transferred to the
mass term by a chiral rotation 
$\psi(x) \rightarrow e^{-i\gamma_5\theta/2} \psi(x)$.
This also means that in the chiral limit, $m \rightarrow m_c$,
the action does not depend on $\theta$.

The flavour singlet axial WI can be derived by making
a local change of variables $\psi(x) \rightarrow 
e^{-i\gamma_5\alpha(x)} \psi(x)$ in the fermionic path integral and
using that the measure is invariant with respect to this.
The variation of the different terms in Eq.\ (\ref{eq:Daction})
give rise to the following terms in the WI,
\begin{equation}
 \partial_\mu j_\mu^5(x) - 2N_f \, q(x) + 
 2(m-m_c)N_f\, \overline{\psi}\gamma_5\psi(x) = 0
    \label{eq:WI}
\end{equation}
By integrating this over all space-time, the divergence of the
axial current does not give any contribution and we arrive
at 
\begin{equation}
  {\cal Q} = -(m-m_c) \mbox{Tr}\left[\gamma_5 D(m)^{-1} \right],
\end{equation}
where ${\cal Q}$ is the integral of $q(x)$ over space-time.
The term ${\cal Q}$ that originates from the Wilson term in the 
action, is usually identified as the topological charge
\cite{Karsten:1980wd}. We would like to emphasise, however
that this identification is based on perturbative arguments that are 
valid only in the naive continuum limit and might also hold in
the proper continuum limit but there is no reason to
believe that any charge obtained directly from gauge field 
observables (e.g.\ by cooling) is a reasonably good approximation
to ${\cal Q}$ in practical situations.

\begin{figure}[htb!]
\centering
\begin{minipage}{8cm}
\resizebox{\textwidth}{!}{
\includegraphics{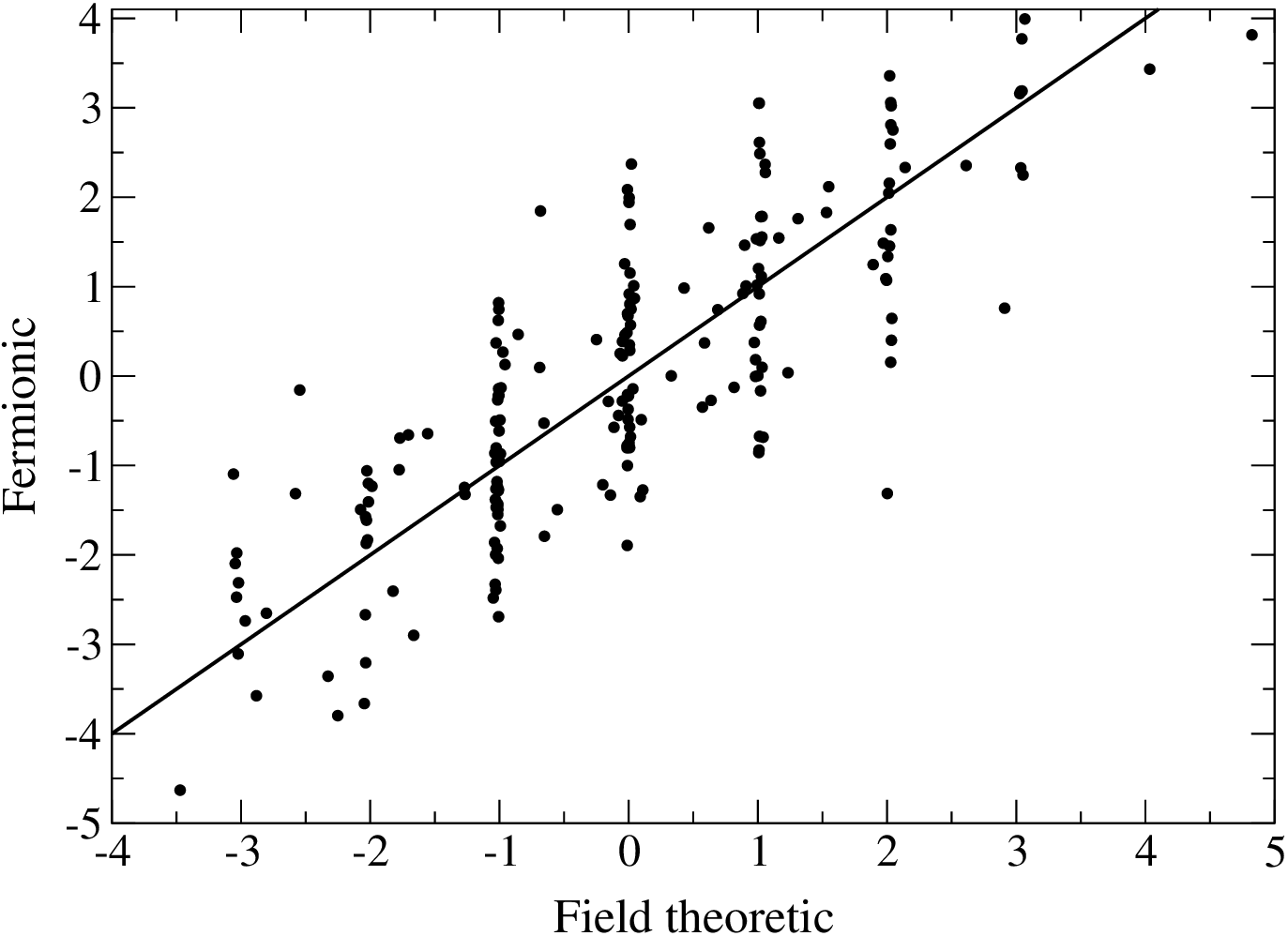}}
\caption{ The value of the 
field theoretic vs.\ the fermionic charge on a set of $\beta=5.85$
quenched gauge configurations.}\label{top_ffd_g5}
\end{minipage}
\end{figure}

In fact, there is good reason to believe that at 
current values of the lattice spacing, these quantities are 
quite different. In Fig.\ \ref{top_ffd_g5} we show a particular
field theoretic charge \cite{Hasenfratz:1998qk} versus ${\cal Q}$
on a set of $\beta=5.85$ quenched configurations\footnote{The ${\cal Q}$
values have been slightly rescaled by an overall factor to give
the same topological susceptibility as the field theoretic charge.}.
Even though, there is a strong correlation between ${\cal Q}$ and
the algebraic charge, $Q$, it is not unlikely that the use
of $Q$ instead of the ``fermionic'' definition ${\cal Q}$
can introduce substantial errors.

A possible way to avoid this would be to use the fermionic
definition, with the Dirac operator 
that enters the dynamical simulation.
There is even an efficient algorithm to 
calculate ${\cal Q}$ \cite{Neff:2001zr}
but difficulties emerging in the chiral limit have
so far made this approach impracticable.

Recently, through the overlap formulation, 
it has become possible to put fermions on the lattice while 
preserving an exact chiral symmetry
\cite{Narayanan:1993ss} and avoiding many of the problems
other formulations have in the chiral limit. In particular,
with the overlap, the above derivation of the WI and the fermionic
definition of the charge results in
\begin{equation}
  {\cal Q} = n_- - n_+,
     \label{eq:Q}
\end{equation}
where $n_\pm$ are the number of positive/negative chirality 
zero modes of $D$. This makes it possible to define the 
``topological charge'' in a consistent way avoiding any reference
to an algebraic charge that has no a priori connection to
the Dirac operator. Another advantage of this formulation is that
with chiral fermions the Leutwyler-Smilga relation has been
shown to hold exactly in the chiral limit even at finite
lattice spacing \cite{Chandrasekharan:1998wg}. This makes
chiral fermions extremely suitable for exploring how the
topological susceptibility interpolates between the chiral
Leutwyler-Smilga and the heavy quark regime.

Unfortunately lattice simulations with dynamical overlap 
fermions have so far been prohibitively expensive. In the present
paper we show the results of a qualitative numerical computation of
the susceptibility with overlap fermions. As we shall see,
the quark-mass dependence of the susceptibility can be estimated 
with a suitable reweighting of a quenched ensemble. 
The procedure we propose is to generate an ensemble $\{U_i\}_1^n$ of
quenched gauge field configurations with only the pure 
gauge measure. In principle the topological susceptibility
can then be computed by reweighting each configuration with
the corresponding fermion determinant, 
\begin{equation}
 \chi(m) = \frac{1}{V \, Z} 
           \sum_i {\cal Q}_i^2 \, \mbox{det}^{N_f}[D(U_i)+m],
     \label{eq:reweight}
\end{equation}
where $D(U_i)$ is the Dirac operator and ${\cal Q}_i$ is the charge
in the background of the gauge configuration $U_i$.
$Z$ is the partition function, i.e.\ the sum appearing in
Eq.\ (\ref{eq:reweight}) without the 
${\cal Q}_i^2$ factor. 

In principle this is a correct procedure,
in practice however it is useless since control over the
statistical errors is lost exponentially with increasing 
volume. The crucial observation is that it is only the small
eigenvalues of $D$ that are correlated with the topological 
charge and without any loss, the full determinant can be 
replaced with a truncation thereof,
\begin{equation}
   \det(D+m) \rightarrow \prod_{k=1}^{N} (\lambda_k + m),
\end{equation}
where $\{\lambda_1,\lambda_2,\dots \lambda_N\}$ are the
$N$ smallest magnitude eigenvalues of $D$.

\begin{figure}[htb!]
\centering
\begin{minipage}{8cm}
\resizebox{\textwidth}{!}{
\includegraphics{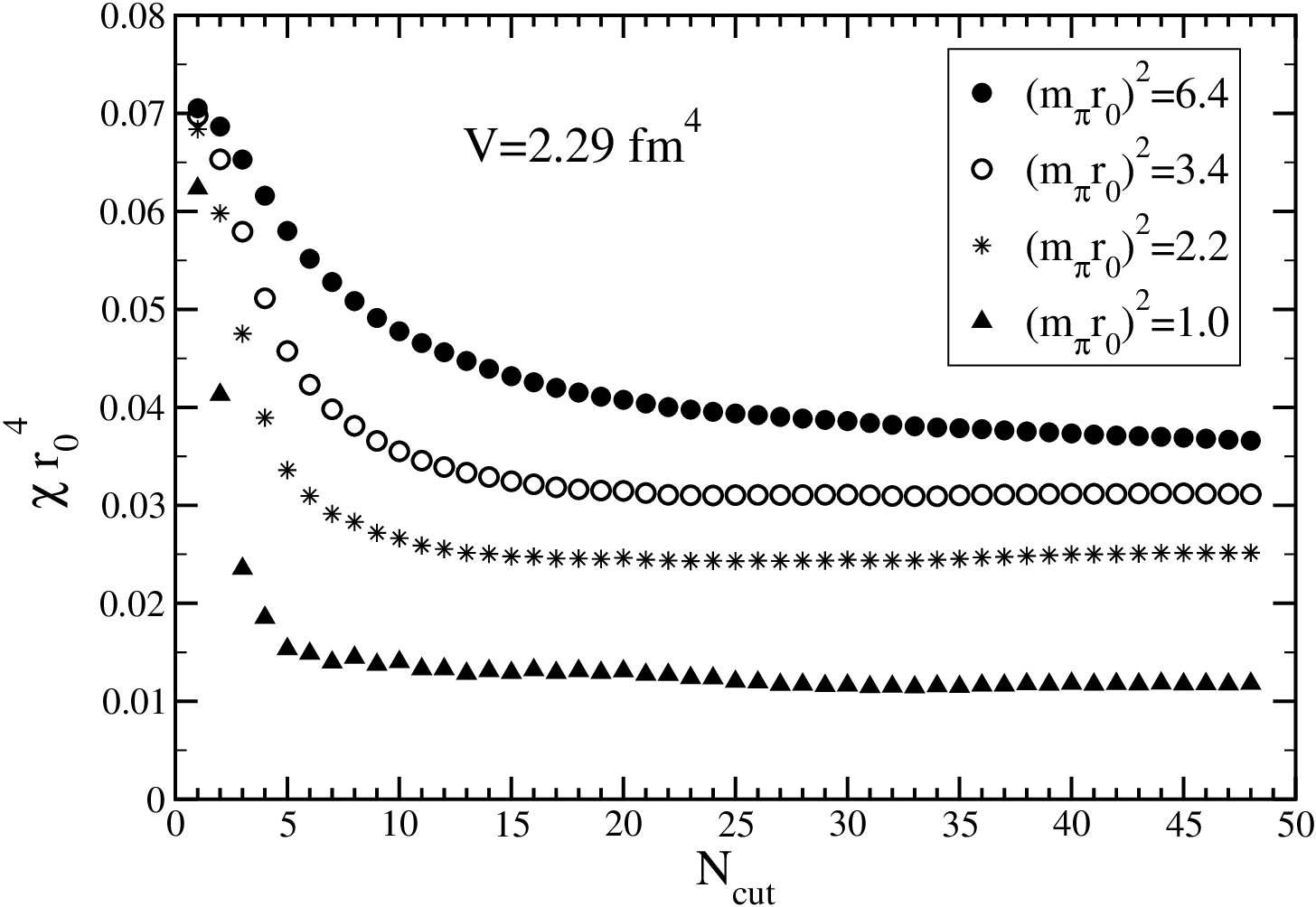}}
\caption{ The reweighted topological  
susceptibility in a $V=2.29$ fm$^4$ box, as a function of the 
number of eigenvalues included in the truncated determinant. The
different curves correspond to different quark (and pion) masses.}
\label{fig:Nincl10}
\end{minipage}
\end{figure}

In Fig.\ \ref{fig:Nincl10}
we show the (reweighted) susceptibility as a function of the
number of eigenvalues included in the truncation. For heavier
quarks, more eigenvalues should be included in the determinant
but overall the truncation is clearly a good approximation.
It drastically reduces the variation of the determinant 
within each topological sector and turns out to make the reweighting 
practicable in a physically useful range of volumes. The upshot is
that once the lowest $N$ eigenvalues of $D$ are known 
on an ensemble of quenched gauge configurations, the full
quark-mass and $N_f$ dependence of the susceptibility can
be easily obtained. The only caveat is that for smaller quark 
mass and larger $N_f$ the determinant fluctuates 
more and consequently the errors increase.

We have seen that the truncation of the fermion determinant 
is a good approximation for the susceptibility but what does
it physically mean to truncate the determinant? We would
like to argue that it is essentially just another possible
action that differs from the untruncated fermion 
action only by local gauge terms. Indeed, it has been shown that
the contribution of the upper part of the spectrum to the
fermion action can be well approximated with a linear combination 
of small Wilson loops \cite{Duncan:1999xh}, i.e.\ a local gauge action.

Besides the susceptibility, there are two other quantities
of interest in our simulation, the lattice spacing (as set by the
Sommer scale $r_0$) and the dependence of the pion mass 
on the bare quark mass. It is expected that including the
truncated determinant will not substantially change $r_0$
compared to its quenched value, in conventional dynamical simulations
most of the change in $r_0$ 
originates from the bulk of the upper part of the
Dirac spectrum. Therefore, in this qualitative calculation
we completely ignore this effect and use the quenched scale.
The change in $m_\pi(m_q)$ compared to the quenched one
can be estimated using the GMOR and the Banks-Casher relation
together with the available low-end of the Dirac spectrum.
The setting of the scale and the pion mass are the two weak 
points of this calculation which might contain uncontrolled
effects. This is the reason why this computation should be
considered only as a qualitative first estimate and hopefully
in the future it will be checked by 
a full-blown dynamical overlap calculation.
We believe, however that even with these caveats, our calculation
is still ``competitive'' with the currently available dynamical calculations
that might have other limitations, as we discussed above.

Simulations at several $\beta$ values and lattice sizes are currently
underway. As an illustration we show results on an ensemble of 
2500 $\beta=5.85$ ($a=0.123$ fm) lattices with a volume of
$V=2.29$ fm$^4$.  Our overlap action
was constructed by inserting the $c_{sw}=1.0$
clover operator with 10 times APE smeared gauge links \cite{DeGrand:1998ss}
into the overlap formula,
\begin{equation}
 D_{ov} = 1 - (1-D_c) 
                   \left[(1-D_c)(1-D_c)^\dagger\right]^{-\frac{1}{2}}
\end{equation}
The inverse square root was approximated with Chebyshev polynomials
after treating the lowest eight eigenmodes of $(1-D_c)(1-D_c)^\dagger$
exactly \cite{Hernandez:2000sb}.

\begin{figure}[htb!]
\centering
\begin{minipage}{9cm}
\resizebox{\textwidth}{!}{
\includegraphics{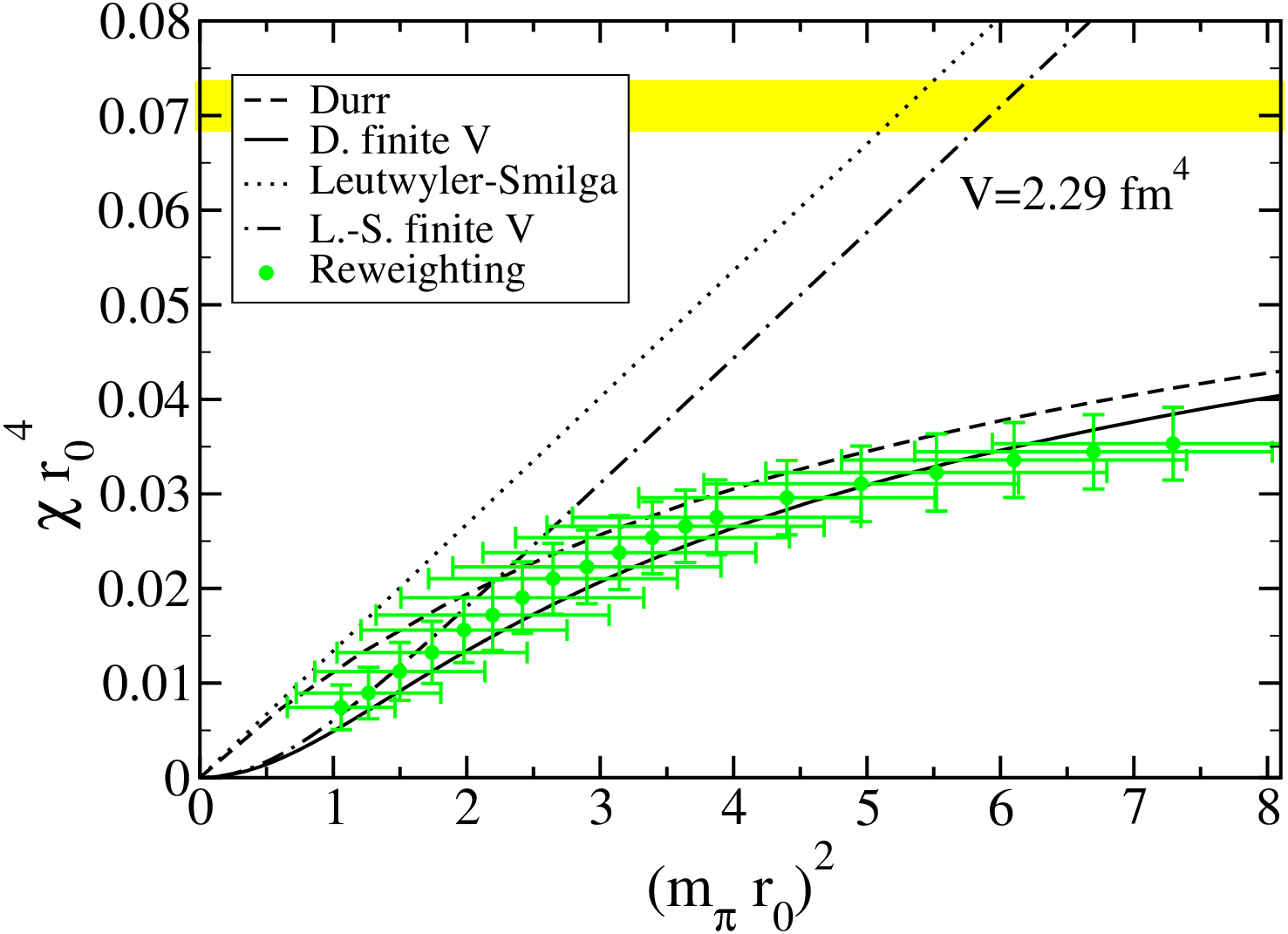}}
\caption{ The topological 
susceptibility vs.\ pion mass in a box volume of $V=2.29$ fm$^4$.}
\label{fig:Nf2ov10}
\end{minipage}
\end{figure}

In Fig.\ \ref{fig:Nf2ov10} we present 
the main result of this paper, the susceptibility versus the 
pion mass with $N_f=2$ flavours of overlap fermions. 
Both the susceptibility and the pion mass are translated into 
dimensionless units with the help of the Sommer scale $r_0=0.49$ fm
\cite{Sommer:1993ce}. The shaded region indicates
the quenched susceptibility $r_0^4 \chi_q = 0.0703(19)$.
obtained on the same ensemble without any reweighting. 
The quoted error is only statistical, we did not include any
additional uncertainty coming from the scale setting.
In more conventional physical units this translates into
$\chi_q = (207.0(1.4) \mbox{MeV})^4$.

For comparison we also plotted the Leutwyler-Smilga formula and
D\"urr's formula of Eq.\ (\ref{eq:Durr})
using our measured quenched susceptibility 
and the known value $f_\pi = 93$ MeV. 
Since both Eq.\ (\ref{eq:Durr}) and (\ref{eq:LS}) are 
valid in the infinite volume limit and our volume is not particularly 
big, we also plotted the curves corresponding to our volume. 

In conclusion, we argued that current computations of the 
topological susceptibility with dynamical quarks might
suffer from a systematic error essentially due to the 
lack of the index theorem at currently used lattice spacings.
We presented a pilot study of the unquenched 
topological susceptibility with $N_f=2$ flavours of overlap
fermions using reweighting. Our computation involved some 
uncontrolled approximations for setting the scale and the pion 
mass which eventually have to be checked in a full-fledged dynamical 
simulation. Nevertheless, our results appear to be consistent with D\"urr's
formula and indicate that the susceptibility might be more 
strongly suppressed than other lattice simulations suggest.
It would also be interesting to compute the susceptibility 
with non-chiral Wilson fermions using the corresponding fermionic
definition of the charge.

\end{document}